\documentclass[aps,prd,superscriptaddress,preprint,nofootinbib]{revtex4}
\usepackage{float}
\usepackage{amsmath,amssymb}%,graphicx}
\usepackage[dvips]{graphicx}
\DeclareGraphicsExtensions{.jpg,.mps,.pdf,.png,.eps} 
\usepackage{color}

\newcommand{\beq}{\begin{equation}}
\newcommand{\eeq}{\end{equation}}

\newcommand{\be}{\begin{equation}}
\newcommand{\ee}{\end{equation}}

\newcommand{\bea}{\begin{eqnarray}}
\newcommand{\eea}{\end{eqnarray}}
\newcommand{\non}{\nonumber}

\begin{document}

\title{ Inflationary quasi-scale-invariant attractors}

\author{Massimiliano Rinaldi}
\email{massimiliano.rinaldi@unitn.it}
\affiliation{Department of Physics, University of Trento\\ Via Sommarive 14, 38123 Trento, Italy}
\affiliation{INFN - TIFPA \\ Via Sommarive 14, 38123 Trento, Italy }

\author{Luciano Vanzo}
\email{luciano.vanzo@unitn.it}
\affiliation{Department of Physics, University of Trento\\ Via Sommarive 14, 38123 Trento, Italy}
\affiliation{INFN - TIFPA \\ Via Sommarive 14, 38123 Trento, Italy }

\author{Sergio Zerbini}
\email{sergio.zerbini@unitn.it}
\affiliation{Department of Physics, University of Trento\\ Via Sommarive 14, 38123 Trento, Italy}
\affiliation{INFN - TIFPA \\ Via Sommarive 14, 38123 Trento, Italy }

\author{Giovanni Venturi}
\email{giovanni.venturi@bo.infn.it}
\affiliation{Dipartimento di Fisica e Astronomia, Universit\`a di Bologna and INFN,\\ Via Irnerio 46, 40126 Bologna, Italy.}

%%%%%%%%%%%%%%%%%%%  ABSTRACT %%%%%%%%%%%%%%%

\begin{abstract} 
\noindent In a series of recent papers Kallosh, Linde, and collaborators have provided a unified description of single-field inflation with several types of potentials, ranging from power law to supergravity, in terms of just one parameter $\alpha$. These so-called $\alpha$-attractors predict a spectral index $n_{s}$ and a tensor-to-scalar ratio $r$, which are fully compatible with the latest Planck data. 
The only common feature of all $\alpha$-attractors is a non-canonical kinetic term with a pole, and a potential analytic around the pole.
In this paper, starting from the same  Einstein frame with a non-canonical scalar kinetic energy,  we explore the case of non-analytic potentials. We  find the functional form that corresponds to quasi-scale invariant gravitational models in the Jordan frame, characterised by a universal relation between $r$ and $n_{s}$ that  fits the observational data but is clearly distinct from the one of the $\alpha$-attractors. It is known that the breaking of the exact classical scale-invariance in the Jordan frame can be attributed to one-loop corrections. Therefore we conclude that there exists a class of non-analytic potentials in the non-canonical Einstein frame that are physically equivalent to a class of  models in the Jordan frame, with  scale-invariance softly broken by one-loop quantum corrections. 
\end{abstract}

 %%%%%%%%%%%%%%%%%%%%%%%%%%%%%%%%%%%%%%%%
 
 \maketitle

%%%%%%%%%%%% INTRO  SECTION %%%%%%%%%%%%%%%%%%%%%

\section{Introduction}

\noindent The most recent data from Planck and BICEP2  have constrained the value of the scalar spectral index $n_{s}$ to unprecedented precision and reduced the upper value of the tensor-to-scalar ratio to about $r<0.10$ \cite{data}. From the physical point of view, this means that the single-field, chaotic inflationary models are greatly favoured  over other competitors. Among these, the ones with potentials of the form  $V(\phi)\sim \phi^{n}$, with $n<2$ provide the best fit, together with the Starobinski model \cite{staro}. The latter can be turned into the Higgs inflationary model \cite{higgs} (at least at high energy) by a conformal transformation, as discussed in
\cite{Kehagias:2013mya}. 

This fact is not merely a coincidence but it rather reflects a universal description of single-field chaotic inflationary models, which also encompasses  the ones descending from supergravity. Such a unified view started in \cite{Ferrara1,Ferrara2}, and explicitely worked out  in \cite{Kallosh1,Kallosh2,Kallosh3}, where the inflaton potential is essentially arbitrary and only assumed to be positive and grow in the vicinity of the moduli space boundary with a stable or metastable minimum. Later, in a series of papers Kallosh, Linde and collaborators have investigated a one-parameter class of attractors, called $\alpha$-attractors, that contains virtually all power-law inflationary models \cite{linde14}. With regard to these $\alpha$-attractors, each member of the class is labeled by the real parameter $\alpha$, which enters the prediction for the spectral indices, given by
\bea
1-n_{s}={2\over N}\,,\quad r={12\alpha\over N^{2}}\,,
\eea
at the leading order in the number of e-foldings $N$. For example, $\alpha=1$ corresponds to the Starobinski model, while, for $\alpha$ large, one recovers the chaotic model with $V\sim \phi^{2}$ \cite{chaot}. Formally,  $\alpha$ can be seen as a deformation parameter in the Lagrangian written in the  Einstein frame  with a non-canonical kinetic term (from now on we call it non-canonical Einstein frame for short) given by
\bea
L=\sqrt{g}\left[\frac{M^2}{2} R-{36\alpha^{2}(\partial\phi)^{2}\over \left(6\alpha-{\phi^{2}}\right)^{2}}-f^{2}\left(\phi\over \sqrt{6\alpha}\right)   \right]\,,
\eea
where $f$ is an arbitrary but \emph{analytic} function near the pole of the kinetic term \cite{kall}. In the Einstein frame with a canonical kinetic term, this amounts to having a plateau in the effective potential, sufficiently flat to accommodate an inflationary evolution.

In this paper we explore the inflationary predictions of a class of non-analytic potentials, in particular with power-law and logarithmic singularities. As we shall see, in this case we find a new class of attractors that fits the data and predicts the relation 
\bea\label{univ1}
r=\frac{8}{3}\left(1-n_s\right)\,,
\eea
at the leading order.

In the case of the $\alpha$-attractors, the fundamental origin of the parameter $\alpha$ is related to supergravity and  provides sound phenomenological predictions. In our case, the common origin of this class of attractors is quite different as  becomes apparent  in the Jordan frame. Indeed, here the Lagrangian turns out to be a deformation of a scale-invariant gravity theory,  which is known to be induced by one-loop quantum corrections. 

In summary, we show that, owing to the analyticity of the scalar potential in the non-canonical Einstein frame,  there can be two inequivalent physical classes of (effective) Lagrangians. The first corresponds to analytic potentials and  the $\alpha$-attractors. The second encompasses a specific form of non-analytic potentials and corresponds to what we call quasi-scale invariant attractors. These two classes are distinct and predict different values for $n_{s}$ and $r$: only more accurate measurements of $r$ will be able to discard one of them.

The paper is organised as follows: in sec.\ \ref{formalism} we lay down the formalism that will be used, and, in particular, we compute the appropriate slow-roll parameters for non-canonical scalar kinetic   term in the Einstein frame. In sec.\ \ref{analytic} we discuss in detail the known case of analytic potentials, along the lines of \cite{linde14}. In sec.\ \ref{nonanalytic} we consider potentials with logarithmic singularities and show, in sec.\ \ref{qsinv}, how these are related to quasi scale-invariant gravity. We conclude in sec.\  \ref{concl} with some considerations.

%%%%%%%%%%% SEC 2 %%%%%%%

\section{Slow-roll parameters for actions with non-canonical scalar kinetic term}\label{formalism}

\noindent Our starting point is the Lagrangian in the non-canonical  Einstein frame  used in \cite{linde14} that we write in the form 
\bea
L=\sqrt{-g}\left( \frac{M^2}{2} R - \frac{A_p}{2\phi^p}(\partial \phi)^2- V(\phi)\right)\,.    
\label{kl}
\eea
Here $M$ is the Planck mass and the Einstein gravitational term is the usual one. The scalar Lagrangian for the self interacting scalar $\phi$  instead  consists of the usual kinetic term multiplied by a  function with a pole of order $p$ in $\phi$ at $\phi=0$, while $A_{p}$ is an arbitrary constant. Of course, one should remember that the presence of this pole singularity is field redefinition dependent, although the one which makes the kinetic term canonical is also singular in general, and in this sense the pole is intrinsic. Nonetheless a non canonical term in the Einstein frame is still very useful for the discussion of  the slow-roll approximation.

In this approximation ($\ddot \phi\simeq \dot\phi^{2}\simeq \dot H\simeq 0$), the equations of motion corresponding to the above Lagrangian,  read
\bea\label{F}
3 M^2 H^2 &\simeq& V(\phi)\,,\\\non\\\label{p}
 \frac{3H^2A_p}{\phi^p} \frac{d \phi }{d N}&\simeq& -V_{\phi}\,,
\eea
where $V_{\phi}={dV\over d\phi}$, and    $N$ is the e-fold time, defined by $dN=Hdt$.  From \eqref{F} and \eqref{p} it follows that 
\bea
N=-\frac{A_{p}}{M^2}\int {Vd\phi\over \phi^{p}V_{\phi}}\,.
\label{N}
\eea
The number of e-foldings between the beginning and the end of inflation, characterized by the corresponding values $\phi_{\rm in}$ and $\phi_{\rm end}$ respectively,  is commonly defined as  
\bea
N_{*}=-\frac{A_{p}}{M^2}\int_{\phi_{\rm end}}^{\phi_{\rm in}} {Vd\phi\over \phi^{p}V_{\phi}}\,.
\eea

\noindent The Hubble flow functions are defined by \cite{ency}
\bea
\epsilon_{0}={H_{0}\over H},\quad \epsilon_{i+1}={\dot \epsilon_{i}\over H\epsilon_{i}},
\eea
where $H_{0}$ is the initial value of the Hubble function. For our work, it is sufficient to consider the first two, which explicitly read 
\bea
\epsilon_{1}&=&-{\dot H\over H^{2}}=-{H'\over H}\,,\\
\epsilon_{2}&=&{\ddot H\over H\dot H}+2\epsilon_{1}={H''\over H'}-{H'\over H}\,,
\eea
where the prime stands for a derivative with respect to $N$. Within the slow-roll approximation, these parameters are related to the observables \cite{Martin:2006rs}. In particular, the spectral index $n_{s}$ and the tensor-to-scalar ratio $r$ are respectively given by
 \bea\label{ns}
n_{s}&\simeq&1-2\epsilon_{1}-\epsilon_{2}, \\\label{ratio}
r&\simeq&16\epsilon_{1}\,.
\eea
In our case, with the help of equations (\ref{F}) and (\ref{p}),  we find that, in the slow-roll regime, the Hubble flow reads
\bea\label{e1}
\epsilon_{1}&\simeq& \frac{M^2}{2}\frac{\phi^p}{A_p}  \left(V_{\phi}\over V\right)^{2},\\\label{e2}
\epsilon_{2}&\simeq& 4\epsilon_{1} - {M^2\phi^{p-1}\over A_{p}}  \left(p{V_{\phi}\over V}+ 2\phi {V_{\phi\phi}\over V} \right)\,.
\eea
Let us note that these expressions for the Hubble flow parameters differ from the standard ones (i.e. the ones computed with a canonical kinetic term in the Lagrangian), which we call $\tilde\epsilon_{1}$ and $\tilde\epsilon_{2}$. The latter are related to the former by $\tilde\epsilon_{1}=\epsilon_{1}(p=0)$ and $\tilde\epsilon_{2}=\epsilon_{2}(p=0)$. 
These formulae are all we need to analyze the classes of attractors corresponding to analytic and non-analytic potentials.

%%%%%%%%%%% SEC 3 %%%%%%%

\section{Analytic potentials}\label{analytic}

\noindent The class of models considered in  \cite{linde14} corresponds to an arbitrary $p >1$ and  analytic potentials at $\phi=0$. Within the slow-roll approximation, we expand $V(\phi)=c_0+c_{1}\phi$, and we assume $c_1/c_{0} >0$ for definitiveness. Thus, from equation \eqref{N} the leading order for small $\phi$ gives, for $p \neq 2$,
\bea
N\simeq{A_{p}c_{0}\phi^{1-p}\over c_{1}M^{2}(p-1)}\,.
\eea
 For $p=2$, one has $N \simeq \frac{1}{\phi}+ O(\ln \phi) $, thus the leading term for small $\phi$ is the same. 

Furthermore, to leading order, one finds that 
\bea
\epsilon_{1}=B_{p}N^{1\over 1-p}\,,
\eea
where 
\bea
B_{p}={{c_{1}^{2}M^{2}K_{p}^{p\over 1-p}\over 2A_{p}c_{0}^{2}}}\,,\quad K_{p}={c_{1}(p-1)M^{2}\over A_{p}c_{0}}\,.
\eea
In addition, one finds that $\epsilon_{1}\ll\epsilon_{2}$ and that, to the leading order
\bea
\epsilon_{2}\simeq {p\over (p-1)N}\,,
\eea
which, in turn, leads to 
\bea
1-n_{s}={p\over (p-1)N}\,,
\eea
together with 
\bea\label{rlinde}
r=16 \epsilon_1=16B_{p} N^{p\over 1-p}\propto (1-n_{s})^{p\over p-1}\,.
\eea
The $p=2$ case is of particular interest, since the Starobinski model \cite{staro}, as well as Higgs inflation \cite{higgs}, belong to this class.   In general, as stressed in \cite{linde14}, the relation between $n_{s}$ and $N$ depends only on $p$.

 %%%%%%%%%%% SEC 4 %%%%%%%

\section{Non-analytic potentials}\label{nonanalytic}

\noindent  Let us now turn our attention to non-analytic potentials. We begin with the simplest models with a singular point at $\phi=0$ represented by 
\bea\label{cond1}
V(\phi,p)=\left\{\begin{array}{ll}a\left(\phi\over \phi_{0}\right) ^{(2-p)/2}\,,  & \quad p > 2\,, \\\\ a\ln \left(\phi\over \phi_{0}\right)\,,  & \quad p=2\,.\end{array}\right.\label{Vln}
\eea
We limit our investigations to these types of singularity because these are typical of potentials with quantum corrections and of interest in inflationary cosmology.  Both forms  satisfy the identity 
\bea
 p V_{\phi}+ 2\phi V_{\phi\phi}=0\,,\quad \forall\quad p\geq 2\,,
\eea
and leads to the relation  (see eq.\ \eqref{e2})
\bea
\epsilon_{2}\simeq 4\epsilon_{1}\,,
\eea
to leading order. Then, from  eqs.\ \eqref{ns} and \eqref{ratio}, we immediately find the relation
\bea
r=\frac{8}{3}\left(1-n_s\right)\,,
\label{univ}
\eea
which is clearly distinct from eq.\ \eqref{rlinde}. From the formula above we have that, for $n_{s}\simeq0.968$, the tensor-to-scalar ratio is fixed to $r\simeq 0.085$, which is a prediction consistent with the latest data and distinct from the one of the $\alpha$-models. The relation \eqref{univ} is also obtained with the addition of an arbitrary constant to the potential: then the first line in Eq.~\eqref{cond1} appears in the context of so called D-brane inflation\cite{Dvali:2001fw}, the second in theories with spontaneously broken SUSY\cite{Dvali:1994ms}, but for the rest of the paper we stick to Eq.~\eqref{cond1} as given. Then for the special case $p=2$, also the number of e-foldings is fixed. On using eq.\ \eqref{e1}, we find that $\epsilon_{1}=1/(4N)$ hence,  $n_{s}=0.968$ implies $N\simeq 47$. For  the  case $p \neq 2$ we find  instead  that
\bea
\epsilon_{1}={1\over 2pN}\,,
\eea
and
\bea
r={8\over pN}\,,\qquad n_{s}=1-{3\over pN}\,.
\eea
We conclude that the class of non-analytic potentials \eqref{Vln} yields the universal relation \eqref{univ}. Each point on this line is  uniquely fixed by the value of the product $pN$, which for the quoted mean value of the tilting by Planck, is within the $68\%$  confidence interval $78.32<Np<114.07$. 

One may wonder what happens when the condition \eqref{cond1} is relaxed. Suppose that the scalar Lagrangian has the form
\bea
L_{\phi}=\sqrt{-g}\left( - \frac{A_p}{2\phi^p}(\partial \phi)^2- {V_{0}\over \phi^{q}}\right)\,,
\eea
where $V_{0}$ is a constant and $q>0$ is not related to $p$. By repeating the calculations above one finds that
\bea
r=\frac83(1-n_{s}){3q\over q+p-2}\,.
\eea
For the special case $p=2$ this relation becomes $r=8(1-n_{s})$ which is excluded by observations. In fact, in order to have an acceptable value of $r$ we need 
\bea
0<{3q\over q+p-2}\leq 1\,.
\eea
With the condition $q>0$ these inequalities yield
\bea
p>2\,,\quad 0<q\leq {p\over 2}-1\,.
\eea
With the help of the e-folding number function \eqref{N} and the definitions of the slow-roll parameters and of the spectral indices, we find that these conditions yield
\bea
r<{4\over N},\quad n_{s}>1-{3\over 2N}\,.
\eea
Therefore, if we fix $N=50$ we find $r<0.08$ but $n_{s}>0.970$. If we increase to $N=60$ we have $r<0.07$ and $n_{s}>0.975$. This means that a non-analytic arbitrary power-law potential does not improve the matching between the theoretical and the observed spectral indices.  Therefore, we continue with potentials of the form \eqref{cond1}, as they seem to be phenomenologically viable and, moreover, appear to have a much deeper interpretation in terms of scale-invariance of the action, as we will discuss in the next section.

%%%%%%%%%%% SEC 5 %%%%%%%

\section{Quasi-scale invariance and non-analytic potential}\label{qsinv}

\noindent To fully understand the fundamental physics behind the results displayed in the last section, it is convenient to go to the Jordan frame (with canonical kinetic term). Let us first consider the $p=2$ case. With the conformal transformation $g_{\mu\nu}\rightarrow A_{2}^{-1}\phi^{2} \,g_{\mu\nu}$, the Lagrangian \eqref{kl} can be written in the convenient form
\bea
L_J=\sqrt{-g}\left[\xi \phi^2 R - \frac{1}{2}(\partial \phi)^2- \lambda \phi^4\left(1+\gamma \ln \frac{\phi^2}{\mu^2}\right)   \right]\,,
\label{gianni}  
\eea
where $\mu$ is an arbitrary mass scale and
\bea
\xi={M^{2}\over 2A_{2}}\,,\quad \gamma=-{1\over 2\ln(\phi_{0}/\mu)}\,,\quad \lambda={V_{0}\ln(\phi_{0}/\mu)\over A_{2}^{4}}\,.
\eea
These three parameters are dimensionless, and the Lagrangian above is exactly scale invariant when $\gamma=0$. When $\gamma\neq 0$, we can interpret this model as a scale-invariant scalar-tensor theory with an  additional logarithmic one-loop quantum correction, which depends on the arbitrary mass scale $\mu^2$, along the lines of the model studied in \cite{gianni}. In general, the logarithmic term  appears in the one-loop expansion of massless scalar quantum electrodynamics, and, in this context, the parameter $\gamma$ depends on the gauge coupling and on the self-interaction scalar coupling. Typically, it is a small number and when it vanishes it simply means that one-loop corrections are negligible. Note that broken scale invariance has also been investigated in references \cite{Finelli,Cerioni,Cerioni1}, and within the context of the so-called Agravity in \cite{Salvio,racio}.
 
It is interesting to note that, on shell and in the slow-roll approximation, this model is equivalent to the deformed quadratic Lagrangian 
\bea\label{onshell}
f(R)=\sqrt{-g}\, \alpha R^2\left(1+\gamma \ln \frac{R}{\mu^2} \right)^{-1}+\cdots\,,
\eea
 proposed in \cite{max1} (for similar models, see \cite{otherR2}), where the relation \eqref{univ} was also found.  To see this, it is sufficient to neglect the kinetic term (slow-roll approximation) and obtain  the equation of motion for $\phi$, which then reads
 \bea\label{phi2}
 \lambda\phi^{2}={\xi R\over 2\left(1+\frac{\gamma}{2}+\gamma\ln \left(\phi^{2}\over \mu^{2}\right)\right)}\,.
 \eea
As shown in \cite{max1}, the solution to this implicit relation is given by the series expansion of the Lambert function \footnote{The Lambert function is defined as the function $z\to W(z)$, solution of $We^{W}=z$.} for large values of the argument, which in the present case has the form
\bea
 \!\!\!\lambda\phi^{2}\!=\!{\xi R \over 2}\left[1+\frac{\gamma}{2}+\gamma\ln \left(R^{2}\over \bar{\mu}^{4}\right)-\lambda\gamma\ln\ln\left(R^{2}\over \bar{\mu}^{4}\right)+\cdots\!\right]^{-1}
 \eea
with $\lambda\gamma\ll\gamma\ll 1$ and a slightly different mass scale $\bar{\mu}$, where the dots are terms of order $\lambda\gamma\ln^{j}\ln R/\ln^{k}R$, $1\leq j< k$. Upon substitution in the Lagrangian \eqref{gianni} we get  the form \eqref{onshell} times a slowly varying  factor (in the variable $R$) given by ratios of logarithmic expressions.

For the case when $p\neq 2$, the conformal transformation is $g_{\mu\nu}\rightarrow A_{p}^{-1}\phi^{p}g_{\mu\nu}$, and  we find
\bea
L_J=\sqrt{-g}\left[\xi_p \phi^{p} R - \frac{1}{2}(\partial \phi)^2- \lambda_p \phi^{2p+{2\over p}-1}   \right]\,,
\label{giannip}  
\eea
where $\xi_{p}=M^{2}/(2A_{p})$ and $\lambda_{p}=V_{0}A_{p}^{-2}\phi_{0}^{(p-2)/p}$ are no longer dimensionless as for $p=2$. Then, we can proceed as before by computing the equation of motion for $\phi$ in the slow-roll approximation, and substituting back into the Lagrangian. As a result, we find that the on-shell expression has the simple form
\bea\label{Rnp}
L\sim \sqrt{g}R^{\,n_{p}}\,,
\eea
where 
\bea\label{nexp}
n_p=2+{2-p\over p^{2}-p+2}\,.
\eea
In \cite{max2}, we  studied the class of $f(R)$ models that best fit the observed values for $n_{s}$ and we found  the simple result
\bea
f(R)=R^{\zeta}\,,\quad \zeta={4-\alpha-\alpha'\over 2(1-\alpha)-\alpha'}\,,
\eea
where $\alpha$ is a function of the e-folding number $N$ and the prime stands for the derivative with respect to $N$. In the limit for constant $\alpha$, the exponent reduces to $\zeta\sim 2+{\cal O}(a)$ thus the Lagrangian again tends to a deformation of $R^{2}$. From \cite{max2} we know that the approximation $\alpha=$ const is not very accurate as it leads to $r\simeq 0.3$ and that, to have a more precise prediction, one needs to consider third-order slow-roll parameters (so that $\alpha'$ is taken in account). Even in this case, one finds that the effective exponent is still very close to 2. This implies that the parameter $p$ cannot be arbitrary if one wants to fit observations. Therefore, we can write  $p=2+\varepsilon$ and the potential in eq.\ \eqref{giannip} can be written, in the limit $\varepsilon\rightarrow 0$, as
\bea
V\simeq \lambda_{p}\phi^{4}\left(1+{3\over 4}\varepsilon\ln {\phi^{2}\over\mu^{2}}\right)\,,
\eea
where, again, $\mu^{2}$ is an arbitrary mass scale to make the argument of the $\log$ dimensionless and where $\varepsilon$ plays the role of the parameter $\gamma$ in eq.\ \eqref{gianni}.

In summary, our main result  is that, to leading order in the slow-roll parameters,  non-analytic potentials of the form \eqref{cond1} in the Einstein frame with non-canonical kinetic term, correspond to theories  in the Jordan frame, represented either by eq.\ \eqref{onshell} or, equivalently, by eq.\ \eqref{gianni}, with scale-invariance broken either by hand or, more realistically, by quantum effects. These models predict the specific relation \eqref{univ} between $r$ and $n_{s}$ and form a set complementary to the $\alpha$-attractors. For this reason, we named these models quasi scale-invariant attractors.

%%%%%%%%%%% SEC 6 %%%%%%%%%

\section{Generalized non-analytic potentials}

\noindent  In this section we generalise the form \eqref{cond1} of the potential to see how robust our results are.  We consider a potential of the form
\bea\label{genpot}
V=V_{0}\left(\phi\over \phi_{0}\right)^{{(2-p)/2}}\sum_{q}a_{q}\left[\ln\left( {\phi\over m}\right)\right]^{q}\,,
\eea
where the logarithmic terms are generically expected from loop quantum corrections, but may also be understood as defining a new class of classical potentials. Here the $a_{q}$ are real coefficients and $m$ is the mass scale beyond which corrections become relevant. Let us now  determine to what extent the relation between $n_{s}$ and $r$ is affected by these (possibly quantum) corrections. For simplicity, we study the potential 
\bea\label{proto}
V=V_{0}\left(\phi\over \phi_{0}\right)^{{(2-p)/2}}\left[1+a\ln\left( {\phi\over m}\right)\right]\,,
\eea
which is sufficient to illustrate the main effect. On using the formalism explained in Sec.\ \ref{formalism}, we compute $\epsilon_{1}$ and $\epsilon_{2}$ and Taylor expand to first order  around $a=0$. We then use these expressions to compute $r$ and $n_{s}$, which depend on $a$ and on $\phi$. Finally, we combine the results to eliminate the $\phi$-dependence, and  find
\bea
r\simeq {8\over 3}(1-n_{s})-{32(1-n_{s})a\over 9(p-2)}+{\cal O}(a^{2})\,,
\eea
which again yields  eq.\ \eqref{univ} in the $a\rightarrow 0$ limit. For $p=2$ this formula does not apply. In fact, in this case we fall back to the case of the simple logarithmic potential studied in the previous section. In general, for $p\neq 0$, we conclude that also with (small) quantum corrections our prediction is  robust: to leading order, logarithmic and power-law non-analytic potentials such as  \eqref{proto} yield the universal relation above. Moreover, positive values of $a/(p-2)$ will tend to further reduce the ratio $r$,  the converse is true  for negative values.

Of course,  these potentials also belong to the class of quasi-scale-invariant Lagrangians in the Jordan frame. In fact, in the slow-roll approximation, we find that, on shell and in the Jordan frame, the relation between $\phi^{2}$ and $R$ is similar to eq.\ \eqref{phi2} and reads
\bea
\lambda \phi^{{p^{2}-p+2}\over p}=\xi R \left[B_{1}+B_{2}\ln\left(\phi\over m\right)\right]^{-1}\,,
\eea
where $B_{1}$ and $B_{2}$ are coefficients depending on $p$ and $V_{0}$.  Thus, as in the previous section, we can always reduce this Lagrangian to the form \eqref{onshell}.

%%%%%%%%%%% CONCLUSIONS %%%%%%%

\section{Conclusions}\label{concl}

\noindent Scale invariance is a powerful physical symmetry that appears in several frameworks. For example, in the context of black hole physics, the invariance of the action is reflected by the thermodynamical properties of the horizon, revealing a deep connection between gravity and entropy \cite{R2blackholes}. In cosmology,  scale invariance was long ago advocated as a fundamental symmetry for the Harrison-ZelÕdovich-Peebles spectrum with $n_{s}=1$ \cite{hzp}. Only when the first measurements revealed that $n_{s}<1$, was this hypothesis , at least partially, abandoned.

If we instead assume  that scale-invariance rules the beginning of inflation, we need a mechanism to break it in order to be consistent with current observations, quite independently of the specific underlying gravitational theory, see e.g.  \cite{stein,mukhanov}. For example, pure $R^{2}$ gravity  admits only a combination of  de Sitter and radiation-dominated solutions \cite{percacci}, where the first  dynamically dominates in time over the second. However, on considering quantum fluctuations, we can envisage a soft breaking mechanism of  the scale invariance of $R^{2}$ via loop quantum corrections. From a phenomenological point of view, this possibility was considered in \cite{max2} and formulated more precisely in \cite{max1}. 

In the present paper, we have shown that the class of quasi scale-invariant models is much larger than the one considered in \cite{max1}. In fact, all potentials with logarithmic and power-law singularities in the Einstein frame  of the form \eqref{cond1} or \eqref{genpot},  and with a non-canonical kinetic term, correspond, at leading order in the slow-roll parameters, to the quasi scale-invariant models represented by the Lagrangian \eqref{onshell}.

For the leading term, all these models yield the same universal relation between $r$ and $n_{s}$, which is independent of the parameters of the theory, in contrast to the case of $\alpha$-attractors, which represent  a complementary alternative. Since our prediction \eqref{univ} is very precise , our model should be easily falsifiable once a more accurate measure of $r$ is  obtained. 

On the other hand , should future observations confirm the universal relation \eqref{univ}, classical scale invariant gravity with loop corrections would become a strong candidate to describe the inflationary Universe in the physical Jordan frame, thus recovering the original spirit of the Harrison-Zeldovich-Peebles model.

\acknowledgments
\noindent We wish to thank G.\ Cognola, M.\ Cicoli, and A.\ Tronconi for valuable discussions. We also  thank A. Linde, R, Kallosh and A. Sagnotti for useful correspondence.

%%%%%%%%%%%%%% BIBLIO %%%%%%%%%%%%%%%%%%%%%%%%%%

\end{document}